\documentclass{emulateapj}

\def\ltsima{$\; \buildrel < \over \sim \;$}
\def\simlt{\lower.5ex\hbox{\ltsima}}
\def\gtsima{$\; \buildrel > \over \sim \;$}
\def\simgt{\lower.5ex\hbox{\gtsima}}

\shorttitle{Two Red Clumps}
\shortauthors{McWilliam \& Zoccali}

\begin{document}

\topmargin 0.5in

\newcommand{\znh}{[{\rm Zn/H}]}
\newcommand{\msol}{M_\odot}
\newcommand{\etal}{et al.\ }
\newcommand{\delv}{\Delta v}
\newcommand{\kms}{km~s$^{-1}$ }
\newcommand{\cm}[1]{\, {\rm cm^{#1}}}
\newcommand{\N}[1]{{N({\rm #1})}}
\newcommand{\e}[1]{{\epsilon({\rm #1})}}
\newcommand{\f}[1]{{f_{\rm #1}}}
\newcommand{\rAA}{{\AA \enskip}}
\newcommand{\sci}[1]{{\rm \; \times \; 10^{#1}}}
\newcommand{\ltk}{\left [ \,}
\newcommand{\ltp}{\left ( \,}
\newcommand{\ltb}{\left \{ \,}
\newcommand{\rtk}{\, \right  ] }
\newcommand{\rtp}{\, \right  ) }
\newcommand{\rtb}{\, \right \} }
\newcommand{\ohf}{{1 \over 2}}
\newcommand{\nohf}{{-1 \over 2}}
\newcommand{\rhf}{{3 \over 2}}
\newcommand{\smm}{\sum\limits}
\newcommand{\perd}{\;\;\; .}
\newcommand{\cmma}{\;\;\; ,}
\newcommand{\intl}{\int\limits}
\newcommand{\mkms}{{\rm \; km\;s^{-1}}}
\newcommand{\ew}{W_\lambda}

\title{Two Red Clumps And The X-Shaped Milky Way Bulge}

\author{Andrew McWilliam}
\affil{The Observatories of the Carnegie Institute of Washington, \\
813 Santa Barbara St., Pasadena, CA 91101--1292}
\email{andy@obs.carnegiescience.edu} 

\and 

\author{Manuela Zoccali}
\affil{Departamento Astronomia y Astrofisica,\\
Pontifcia Universidad Catolica de Chile,\\
Avenida Vicuna Mackenna 4860, 
Santiago, Chile}
\email{mzoccali@astro.puc.cl}

\begin{abstract}
From 2MASS infra-red photometry we find two red clump (RC)
populations co-existing in fields toward the Galactic bulge
at latitudes $|b|>$5.5$^{\circ}$, ranging over $\sim$13$^{\circ}$ in
longitude and 20$^{\circ}$ in latitude.  These RC peaks indicate 
two stellar populations separated by $\sim$2.3 kpc; at
($l,b$)=($+1,-8$) the two RCs are located at 6.5 and 8.8$\pm$0.2
kpc.  The double-peaked RC is inconsistent with a tilted bar morphology.
Most of our fields show the two RCs at roughly
constant distance with longitude, also inconsistent with a tilted bar;
however, an underlying bar may be present.

Stellar densities in the two RCs changes dramatically
with longitude: on the positive longitude side the foreground RC
is dominant, while the background RC dominates negative longitudes.
A line connecting the maxima of the foreground and background
populations is tilted to the line of sight by $\sim$20$\pm4^{\circ}$,
similar to claims for the tilt of a Galactic bar.
The distance between the two RCs decreases towards the Galactic
plane; seen edge-on the bulge is X-shaped, resembling
some extra-galactic bulges and the results of N-body simulations.
The center of this X is consistent with the distance
to the Galactic center, although better agreement would occur if
the bulge is 2--3 Gyr younger than 47~Tuc.

Our observations may be understood if the two RC populations
emanate, nearly tangentially, from the Galactic bar
ends, in a funnel shape.  Alternatively, the X, or double funnel,
may continue to the Galactic center.  From the Sun this would appear 
peanut/box shaped, but X-shaped when viewed tangentially.
\end{abstract}


\section{Introduction}

Following the detection of an apparent H~I bar within 2kpc of the Galactic
center, by Liszt \& Burton (1980), Blitz \& Spergel (1991) found that the
2.4$\mu$ imaging data of Matsumoto et al. (1982) showed the presence of a tilted
bar in the Galactic bulge.  That part of the bar closest to the sun was
at positive Galactic longitudes.  These conclusions were primarily based on the
observed fluxes for latitudes $|b|$=3--9$^{\circ}$ and longitudes at 
$l$=0,$\pm$10$^{\circ}$.

Stanek et al. (1994, 1997) studied V and I photometry of Red Clump (RC) stars
toward the bulge at longitudes of 
$l$$\sim$1,$\pm$5$^{\circ}$ (latitudes near $b$=$-$4$^{\circ}$) and
found a systematically fainter RCs from positive to negative longitude.  
This was interpreted, and modelled, as a distance effect which was fit
with a tri-axial structure, or tilted bar, with the near-side at positive 
Galactic longitudes.

Later work on the COBE/DIRBE near and far infrared sky brightnesses data by
Dwek et al. (1995) and Binney, Gerhard, \& Spergel (1997) confirmed
the existence of the tilted Galactic bar.  Distances from RC stars,
studied for fields at various positions, from Babusiaux and Gilmore (2005),
Nishiyama et al. (2005, 2006), and Rattenbury et al. (2007) confirm
the general picture of a tilted Galactic bar.  In particular, Nishiyama
et al. (2005, 2006) claimed to find a secondary inner bar.

The recent radial velocity study of bulge M giants stars, by Howard et al.
(2008, 2009), found rotation that is best fit with bar models. In particular,
they find that the bulge at $-$8$^\circ$ and $-$4$^\circ$ rotates cylindrically, 
as do boxy bulges of other galaxies.

Recently, McWilliam et al. (2010) and Zoccali (2010) independently found
evidence for two red clumps toward the Galactic bulge region.  In the current paper
we outline and expand on the evidence of this joint discovery and map the
extent of the double red clump over the bulge region.

\section{Observational Data}

\begin{figure*}[ht]
\centering
\includegraphics[angle=-90,width=15cm]{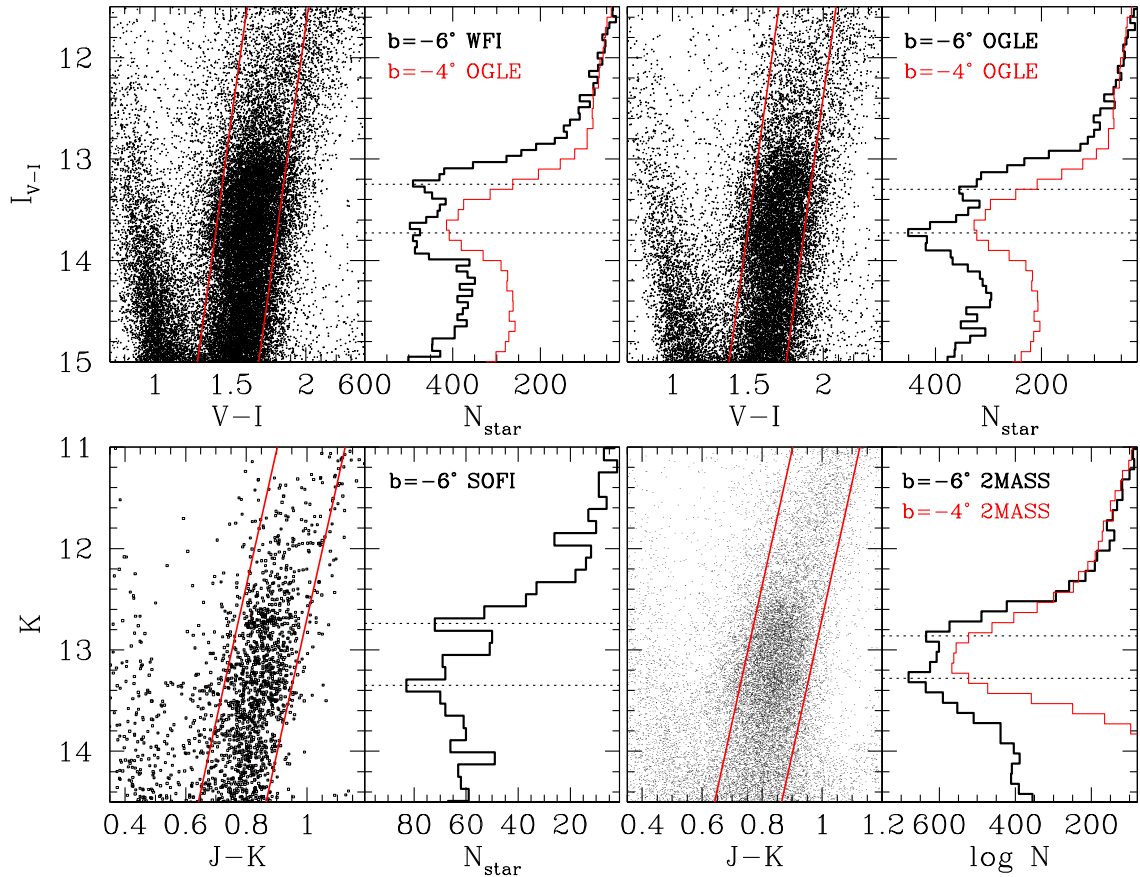}
\caption{ CMDs and  LFs for the  red clump
regions  in   the  field  at  $(l,b)=(0,-6)$   from  four  independent
photometric catalogs. The two panels in the upper left side show the
WFI  photometry, at the  upper right  is the  OGLE photometry,  at the
lower left  the SOFI photometry and  at the lower right  the 2MASS one
(see  text for references).  A diagonal  strip in  the CMDs  shows the
region selected to construct the histograms. A thin histogram show the
red clump region in the Baade's Window field, at $(l,b)=(0,-4)$. }
\label{cmd_b6}
\end{figure*}

The data discussed here come from the following four photometric catalogs
available in the literature:

{\it i)} The 2MASS point source catalog (Skrutskie et al. 2006).

{\it ii)} Optical  $V,I$  WFI photometry  at  the 2.2m  ESO/MPG
telescope   at  La  Silla,   for  a   $34\times33$  arcmin   field  at
$(l,b)=(0,-5.4)$ (Zoccali et al. 2003).

{\it iii)} Near  IR $J,H,K$  photometry from  SOFI at  the NTT  in La
Silla,    for   a    $8.3\times8.3$   arcmin    field    centered   at
$(l,b)=(0.28,-6.17)$ (Zoccali et al. 2003).

{\it iv)} The  OGLE $B,V,I$ maps of the  Galactic bulge, from Udalski
et al. (2002).

In Figure~\ref{cmd_b6} we show the  color magnitude diagram (CMD) of the
bulge region  centered approximately at $b=-6^\circ$,  along the minor
axis, from four independent photometric data sets, all of them showing
a double peaked  horizontal branch  red clump.  The upper  panels show  
the WFI (left)  and OGLE  (right) optical  CMDs together  with  the luminosity
functions (LFs) of RGB stars  falling inside the diagonal strip marked
in the CMD.  The reddening  free magnitude $I_{\rm V-I}$ 
\footnote{The    $I_{\rm    V-I}$     magnitude    is    defined    as
$I_{V-I}=I-(V-I) \times A_I/(A_V-A_I)$.}  was used, following  Rattenbury  et
al.  (2007), and  adopting  the  extinction values  given  in Sumi  et
al.  (2004) for  the OGLE  fields. The OGLE field shown here  is the
number  7, centered  at $(l,b)=(-0.14,-5.91)$.   It is  worth noticing
that Rattenbury et  al. (2007) data also show a double  peaked red clump 
in this field (c.f.,  their Fig.~5).  The  red thin  histograms in  both
panels  show the  LF  of the  OGLE  field centered  in Baade's  Window
(SC46), where the red clump is significantly narrower and uni-modal.

The lower  panels of Figure~\ref{cmd_b6} show  the near IR  CMDs and LFs
from the SOFI  photometry (left) and the 2MASS  point source catalog
(right). Again, the  red clump is double peaked in  this field. The LF
for Baade's Window is also shown  as a red thin histogram.  The latter
was obtained  by selecting a 0.8  square degree field,  centered at at
$(l,b)=(0,-4)$, from  the 2MASS point  source catalog. Unfortunately
at $b=-4^\circ$ the completeness of the 2MASS catalog drops abruptly
below the red clump, therefore  the comparison is not as conclusive as
the one for the OGLE catalogs.

\begin{figure}[ht]
\centering
\includegraphics[angle=0,width=9cm]{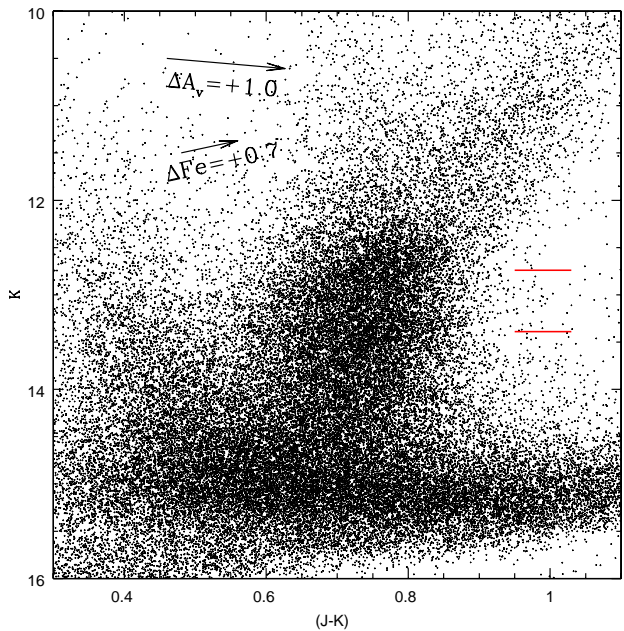}
\caption{The 2MASS K,(J$-$K) color-magnitude diagram for the field at ($l$,$b$)=($-$1,$-$8).
The two red clumps are visible showing a slight upward trend.  Red lines indicate
the observed K-band RC peaks in the Plaut field at ($l$,$b$)=($+$1,$-$8).  The upward
pointing vector shows the change in RC K,(J$-$K) from [Fe/H]=$-$0.70 to 0.00, predicted from
the Teramo stellar evolution models (e.g., Pietrinferni et al. 2004) and is consistent with
the observed RC slopes.  Also shown is a downward reddening vector for 
$\Delta$A$_{\rm v}$=1.0 mag.  }
\label{fig-cmd-10-80}
\end{figure}

The 2MASS color-magnitude diagram in Figure~\ref{fig-cmd-10-80} shows 
the two RCs at ($l,b$)=($-$1,$-$8), which appear tilted, consistent with
the predicted metallicity-dependence of the Teramo isochrones 
(Pietrinferni et al. 2004).  The figure also clearly shows that the
two RCs possess similar range and mean (J$-$K) colors; as in Figure~1 this
suggests that the reddening is foreground and small, and that the two 
RC populations possess a similar range and mean metallicities.

The mean (J$-$K)$_{\rm 2MASS}$ color differences between bright and faint
RCs for the 16 fields in this paper that contain both RCs is 0.017$\pm$0.003
magnitudes.  Thus, the colors are practically identical, but in all cases
the brighter RC is slightly redder than the faint RC.  This systematic
difference is in the right sense and size to be due to the change in color
of the background red giant branch population between the bright and faint
RCs.

A small metallicity difference (about 0.1 dex), between bright and faint RCs
could produce the color shift, but it would still be necessary to subtract
the red giant branch background effect; thus, any metallicity difference
must be smaller than $\sim$0.1 dex.

Figure~\ref{cmd_b6} and \ref{fig-cmd-10-80} suggests that the double clump
might be due to the presence of two populations at two different distances.
A magnitude difference of $\sim 0.4$ at $\sim 8$ kpc would correspond to
a distance difference of $\sim 1.5$ kpc.

We note here for the first time that in the outer bulge, along the minor 
axis, bright and faint red clumps coexist, as if the near and far side 
of the bar both extend  towards  the minor  axis.  

Figure~\ref{cmd_b6} demonstrates that the horizontal branch red clump in
the field at $b=-6^\circ$ is significantly broader than the one of 
Baade's Window at $b=-4^\circ$. 

If the double-peaked RCs are due to the distances of two populations,
the observations immediately appear inconsistent with a single tilted bar.
It is, therefore, important to ask whether the double-peaked RCs could
have resulted from stellar evolution, or to effects other than distance.

A few points can be addressed by looking at these figures. 
First of all, the double clump is real. It is not an artifact of bad
photometry, such as a bad match of mosaic data, because it is present
in several independent catalogs. We have checked that it is present
in each of the 8 chips of the WFI mosaic. 

Second, the two peaks cannot be due to the RGB bump (nor the AGB bump)
falling close to the RC because in that case
the two would also occur in Baade's Window.  

Also, as we will  see in Figures~\ref{fig-lon} \& \ref{fig-lon7}, the
relative  strength of  the two RC peaks changes dramatically with 
longitude, while the  population ratios of RGB bump, AGB  bump and HB red
clump depend  on the  evolutionary time of these  phases, and
therefore should be independent of the line of sight.

The double clump cannot be due to two extinction patches in
this particular direction, because in that case the separation between
the two peaks would be smaller in the $K$ band, compared to the $I$
band.  Instead, the separation is roughly comparable in all the
filters.  In addition, as mentioned above, the two peaks look very
similar in the 8 chips of the WFI mosaic. This would  not be the case
if they were due to extinction  patches.  Finally, were extinction
responsible for the two peaks, the clump that is $\Delta I\sim 0.4$
magnitudes fainter would also be $\sim 0.41$ magnitudes redder than
the other one in $V-I$, which is clearly not the case.

The predicted RC$-$RGB bump I-band magnitude differences from the Teramo stellar evolution
code (e.g., Pietrinferni et al. 2004) increase with metallicity, and a match to the
observed difference between our putative RCs occurs for 12Gyr isochrone at solar
metallicity.  However, the predicted RC numbers exceed those of the RGB bump
by roughly a factor of 10.  This presents a particularly severe problem for the negative
longitude fields (see Figure~\ref{fig-lon}), where the faint component dominates, 
so the proposed RGB bump population exceeds the RC, completely at odds with current
stellar evolution ideas.  The RGB bump/RC ratio for the negative longitude bulge
fields of Figure~\ref{fig-lon} exceed the stellar evolution predictions by more than a
factor of 30.  This might be qualitatively understood 
if metal-rich bulge stars largely terminate their evolution prematurely, after the
RGB bump but before the RC, say if their 
metallicity is higher than currently realized and they experienced significantly
enhanced mass-loss.  However, as noted above, it would then be necessary to understand
why stellar evolution on the giant branch is so different on the positive and negative
longitude sides of the bulge; as a result there would still be two separate bulge
populations.  The similarity of the J$-$K colors of the two populations
suggests that there is very little metallicity difference.
Given these difficulties, the idea that our two RCs are due to RC plus RGB bump
is not a tenable hypothesis.  It is likely that the RGB bump is detected as a
very small peak in some of the panels of Figure~\ref{fig-lon}; the main effect of the
RGB bump is to reduce the clarity of the fainter RC peaks.

Theoretical isochrone RC K-band magnitudes and (J$-$K) colors display
metallicity-dependence: more metal-rich RC stars are redder in (J$-$K)
and brighter in K.  This effect is clearly evident in the Teramo group
stellar evolution models (e.g., Pietrinferni et al. 2004).

The theoretical predictions show that RC M$_K$ is brighter for younger ages;
however, no single age can account for the observed 0.65 mag. difference in RC K$_0$
seen of Figure~\ref{fig-cmd-10-80} and \ref{fig-lon} for solar-metallicity
isochrones, unlike the case for the I-band.  The closest that
the theoretical RC predictions come to to the observations, at solar composition,
is for ages at 2 and 14 Gyrs, but this only accounts for $\Delta$M$_K$ of 0.28 mag.
A combination of age and metallicity effects could explain the brightness of the
observed RC peaks, but this requires unrealistic ages and a mean [Fe/H] near
$-$0.5 dex, which is in conflict with measured bulge abundances.  For latitude
$-$8$^\circ$ the age explanation would require the majority of stars at positive
longitudes to be 2Gyr old and the majority of stars at negative longitudes to be
14 Gyr, with a roughly equal mixture of both at longitudes near $l$$\sim$0.  However,
this is excluded by the age data of Zoccali et al. (2003), at
($l$,$b$)=(0.3,$-$6.2), and Clarkson et al. (2008) at ($l$,$b$)=(1.3,$-$2.7).
Both studies find an old bulge $\geq$10 Gyr with almost no trace of a younger
population.  Again, this would not explain how two
vastly different age stellar populations could be maintained on opposite 
sides of the Galactic center.  We
conclude that it is not possible to account for the observed K-band RC 
magnitudes with two stellar populations of different ages at a single distance.
Thus, we attribute the observed differences in RC magnitude as primarily a distance
effect.

  We note the similarity of our bulge-wide double RC to the double RC in the bulge
  globular cluster Terzan~5, found by Ferraro et al. (2009).  It is 
  tantalizing that the Galactic longitude of Terzan~5 puts it very close to the peak
  of our bright RC (but closer to the plane than our fields), and that Ferraro
  et al.'s preferred Terzan~5 distance, at 5.9$\pm$0.5 kpc, is very similar to
  our estimated distance for the bright RC, at 6.5 $\pm$0.2 kpc.

  Ferraro et al. (2009) fit the RC region using two isochrones, one at 12 Gyr with
  [Fe/H]=$-$0.2 dex, and a brighter one with an age of 6 Gyr and [Fe/H]=$+$0.3 dex.
  The young population is critical for explaining the Terzan~5 double RC.
  If this explanation were true for the bulge-wide double RC found here it might
  indicate that the bright RC is due to a huge accreted stellar system.

  While a second, young, metal-rich, population is acceptable for an individual 
  globular cluster, it is in conflict with the distribution of {\it cmd} ages
  found by Zoccali et al. (2003) and Clarkson et al. (2008), who find almost no
  trace of bulge stars with ages less than 10~Gyr.  The Zoccali et al. (2003) field
  is closest to our fields near $(l,b)$=$(0,-8)$, which shows a large fraction of
  bright and faint RC star populations.  Thus, for double RCs due to age and
  metallicity, rather than distance, there should be large numbers of both young
  and old populations in the Zoccali et al. (2003) field; yet no significant 6~Gyr
  population is evident in the Zoccali et al. (2003) data.

The two RCs in Ter~5 $cmd$ are shifted in mean (J$-$K) color by $\sim$0.15 magnitudes,
consistent with the claimed [Fe/H] difference.
However, for our bulge-wide fields the
  bright and faint RCs have nearly identical mean colors, no matter what the reddening,
  suggesting no significant metallicity differences.  Without a metallicity difference
  the required ages of the two bulge-wide RC populations, for [Fe/H]$\sim$$-$0.2, needs
  to be 2 and 14 Gyrs, as stated above; such age differences are easily ruled-out.  Thus,
  there seems to be no way to appeal to an acceptable combination of age and metallicity
  to explain the brightness and color of the two bulge-wide double RCs
  discussed in this paper; thus, we assume that the double RCs reflect a distance effect.

  A caveat is that this conclusion relies on the assumption that the age distribution
  at $(l,b)$=$(0,-8)$ is similar to the old ages in the $b=-6.2^{\circ}$ field of
  Zoccali et al. (2003) and the $b=-2.7^{\circ}$ field of Clarkson et al. (2008).

\section{The Double Red Clump Across the Bulge}

\subsection{Extent in Longitude}

\begin{figure*}[ht]
\centering
\includegraphics[angle=0,width=15cm]{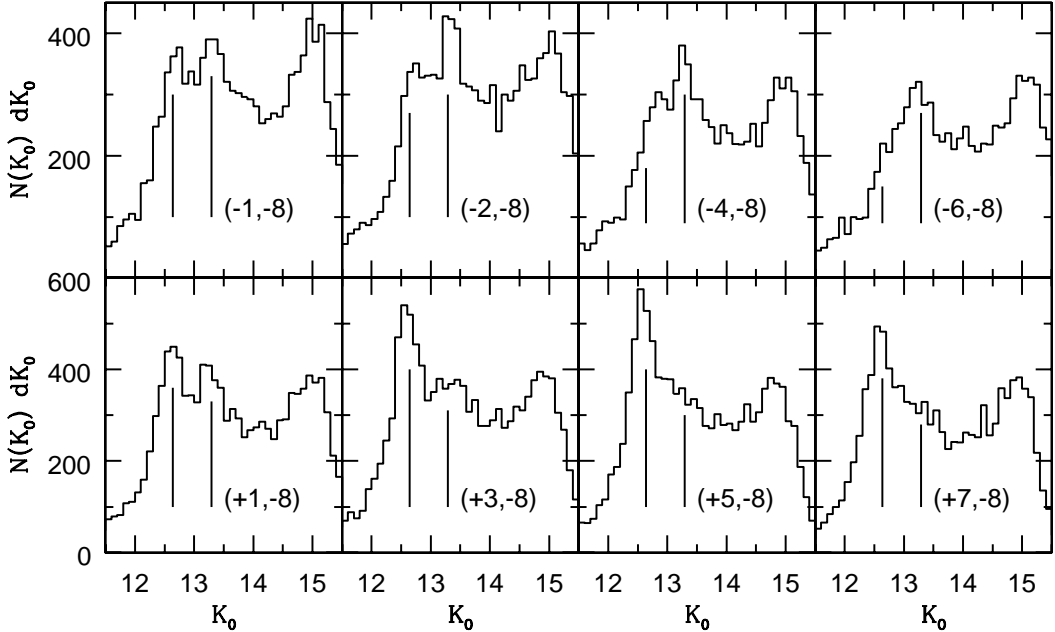}
\caption{Luminosity functions for the red clump region of fields at
various longitudes, for latitude $b=-8^\circ$. The bright red clump
component is particularly strong on the positive longitude side, while
the faint component is stronger on the negative longitude side.}
\label{fig-lon}
\end{figure*}

In order  to map  out the extent  of the  double RC region  toward the
Galactic   bulge  we   have   produced  CMDs   and   LFs  similar   to
Figure~\ref{cmd_b6} for a large number of regions.  In
Figure~\ref{fig-lon}  we show  some  of these regions for  latitude 
$b=-8^\circ$, with  longitudes ranging from  $l=-6$$^\circ$ to $+7$$^\circ$.

The mean reddening of  each field was estimated from the  mean 2MASS (J$-$K) 
color of the red clump.  Here we adopt an unreddened 2MASS (J$-$K)$_0$=0.61
for the $-$8$^{\circ}$ bulge red clump, based on the Alves (2000) mean CIT
V$-$K=2.35 for solar neighborhood RC stars, the Alonso et al. (1999)
color-T$_{\rm eff}$ relations and the CIT--2MASS color transformations of
Carpenter (2005 unpublished)
\footnote {http://www.astro.caltech.edu/~jmc/2mass/v3/transformations}.  
The mean metallicity of the Alves (2000) solar neighborhood RC stars is similar 
to the mean metallicity of the $-$8$^{\circ}$ bulge field, at [Fe/H]=$-$0.18 dex,
based on a linear metallicity gradient computed with the metallicities of
nearby fields from Zoccali et al. (2008). Thus, the Alves (2000)
RC calibration has the appropriate metallicity for this bulge field,
and roughly the right T$_{\rm eff}$ and (J$-$K) color,
although because the  bulge is significantly older the luminosity
of the Solar neighborhood RC stars is different.  Note that because the RC
(J$-$K) colors show a small metallicity sensitivity our reddening values are
close, but not strictly correct, for bulge fields with different mean
metallicities.  In our analysis we employ
the Winkler (1997) reddening law, from which we find A$_K$=0.64 E(J$-$K).
We note that Nishiyama et al (2009) find a slightly different reddening law 
for the bulge, for which they obtain A$_K$=0.528 E(J$-$K).  Our K-band extinction 
correction for ($l$,$b$)=($-$1,$-$8) at 0.10 mag. would be reduced to 0.08 mag.
if we employed the Nishiyama et al. (2009) reddening law.  Because these values
are very small, being less than the bin width in the luminosity functions
presented here, the choice of reddening law does not affect our conclusions.

We prefer to use the (J$-$K) color difference to estimate the average reddening
and K-band extinction for each field rather than the (H$-$K) color because the
bulge RC (H$-$K) color, in our fields, suffers significant overlap with foreground disk
Main Sequence and RC stars.

We do not attempt to correct for extinction of individual RC stars using the
reddening-free K$_{HK}$ magnitudes employed by Nishiyama et al. (2005), because
this introduces unacceptable errors for our fields.  The formula used by
Nishiyama et al. (2005) effectively adopts reddening values for each star based
on the H$-$K color distance from the mean RC value.  However, the RC is
not a discrete point, but has an intrinsic color width even for single populations;
in 47~Tuc the RC has a 2MASS color width of 0.10 magnitudes in (J$-$K) and 0.13 
magnitudes in (H$-$K).  
Because the bulge contains stars with a range of metallicity the intrinsic
width of the RC is even larger than in 47~Tuc.  For this reason, distance
from the mean RC color is a poor reddening indicator in the low-extinction bulge fields
considered here.  If
we employ K$_{HK}$ from Nishiyama et al. (2005) our bulge luminosity functions
would effectively be convolved with an error distribution and the peaks more
difficult to identify.

Because the K-band extinctions of our fields are small, typically less than 0.10
magnitudes, it is more reliable to simply correct for the average extinction
for each field, implied from the mean RC (J$-$K) color, than to employ
the putative reddening-free K$_{HK}$ magnitudes for individual stars.  
Because of the metallicity-dependence of the RC (J$-$K) and M$_K$ values
our adopted mean extinctions and implied distances are sensitive to systematic
metallicity differences between fields.

In Figure~\ref{fig-lon} we show the dereddened 2MASS K-band luminosity
function, for 8 fields covering longitudes from $l$=$-$6 to $l$=$+$7 degrees,
at a latitude of $b$=$-$8 degrees.  Each field was constructed using 2MASS data
from a 30 arc-minute radius circular aperture on the sky.  The luminosity function 
was derived from a vertical strip, 0.10 magnitudes wide in (J$-$K)$_0$, centered
on the mean RC (J$-$K)$_0$ color for each field.  The color-magnitude
diagrams of all 8 fields show that the bright and faint RC possess similar ranges
and mean (J$-$K)$_0$ values, as also seen in Figures 1 and 2.  This is most
easily understood if the two RCs have similar metallicities and reddenings
(i.e., due to foreground reddening).
For fields beyond the ranges of Figure~\ref{fig-lon} we can see the same
RCs individually, but not both together, over a large area of sky
approximately 20$\times$20 degrees in extent.

Globular clusters have a negligible effect on our luminosity functions.
In Figure~\ref{fig-lon} only the field at ($l$,$b$)=$+$3,$-$8 contains
a globular cluster, NGC~6624, which covers an extremely small fraction of the field.
The distance modulus of NGC~6624 (Harris et al. 1996) indicates K$_0$=14.1, which
coincides with a minuscule peak in the luminosity function of 
Figure~\ref{fig-lon}, and is comparable to the noise.

We determine distances of the two RC populations using the observed 47~Tuc RC plus
theoretical corrections for metallicity.
For the bulge field in Figure~\ref{fig-lon} at ($l$,$b$)=($+$1,$-$8) the bright and
faint RC peaks are located at K$_0$=12.64 and 13.29 respectively.  
Zoccali et al. (2008) suggests that [Fe/H]=$-$0.18 for a bulge latitude of $b=-8^{\circ}$;
Koch \& McWilliam (2008) find [Fe/H] for 47~Tuc of $-$0.76 dex.
The Teramo stellar evolution models indicate that the bulge RC is 0.20 magnitudes more
luminous than the 47~Tuc in the K-band, due to this metallicity difference, assuming
that both systems are 12 Gyr old.  If the bulge is younger or more metal-rich the size
of the correction increases.  The 47~Tuc RC 2MASS K$_0$ magnitude is 11.98 and its 
distance modulus is 13.22 (Koch \& McWilliam 2008).  Combined with the metallicity 
K-band correction we find distances for the bright and faint bulge RCs of 6.5 and 
8.8 $\pm$0.2 Kpc respectively.

As noted earlier, the ratio of bright to faint RC in Figure~\ref{fig-lon}
varies strongly with longitude, with the faint RC dominant at negative
longitudes, and the bright peak strongest on the positive longitude
side of the bulge.  For the latitude $b=-8^{\circ}$ fields of Figure~\ref{fig-lon}
the faint RC peaks around $l=-2^{\circ}$ and the bright RC peaks near $l=+5^{\circ}$.

We note that the foreground peak, at $l=+5^{\circ}$ contains many
more RC stars than the peak of the background population, at $l=-2^{\circ}$.   
Because the background population is at a higher distance above the
plane for $b=-8^{\circ}$, a lower star count is expected for the background RC.
This provides a qualitative explanation why the negative longitude side of
the bulge is fainter at NIR wavelengths than at positive longitudes away from
the Galactic plane.

The panels in  Figure~\ref{fig-lon} show that the two  RCs remain
at similar K$_0$ magnitudes, at latitude $b=-8^{\circ}$, although the
relative strength of the faint
and bright RCs changes strongly  with longitude, with the  bright one
dominating  at positive  longitudes,  and the  faint  one at  negative
longitudes.  Closer to  the plane, at  latitudes $|b|<4^{\circ}$,
similar behavior is observed:  a (single) bright clump at positive $l$,
becoming progressively  fainter as one crosses  the minor
axis and  moves towards negative $l$. 

While the RC populations in the panels of Figure~\ref{fig-lon} appear at
nearly constant K$_0$ magnitude, the ($l,b$)=($-4,-8$) and ($+5,-8$)
panels provide the strongest evidence of locations where the foreground
RC is more distant and the background RC closer than the norm.
It would be interesting to know whether these distance changes are consistent 
with spiral locii.

\begin{figure*}[ht]
\centering
\includegraphics[angle=0,width=15cm]{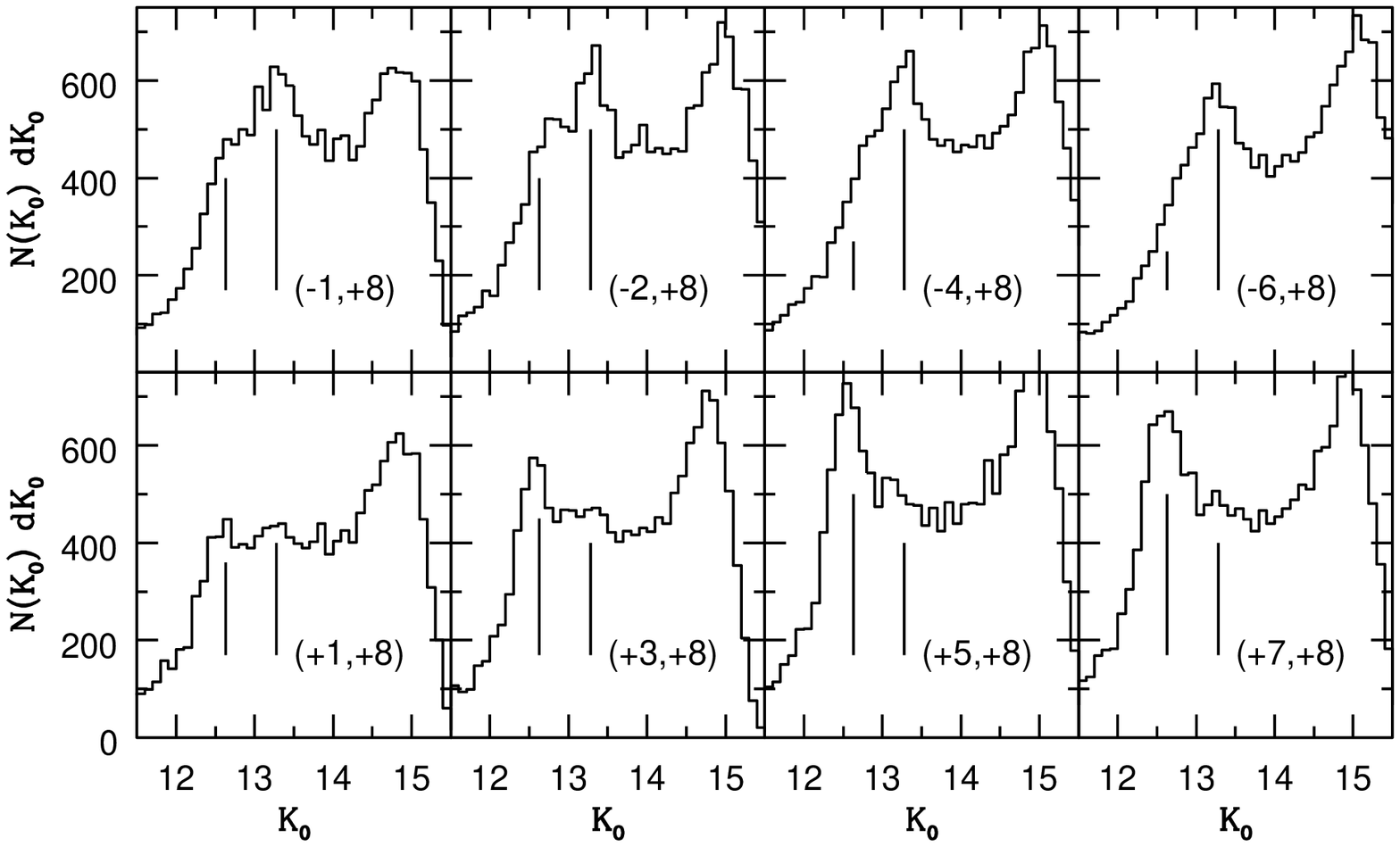}
\caption{Luminosity functions for the red clump region of fields at
various longitudes, for latitude $b=+8^\circ$. The bright red clump
component is particularly strong on the positive longitude side, while
the faint component is stronger on the negative longitude side.}
\label{fig-lon7}
\end{figure*}

The same two-component RC behavior is seen on the positive latitude side,
above the Galactic plane, as evident in Figure~\ref{fig-lon7}.
Compared to the fields at $b=-8^{\circ}$ the double RC peaks are less
pronounced at $b=+8^{\circ}$.  On the positive latitude side the
peak of the RC number counts is near $l=+6$ and $l=-3$ for the foreground and
background populations respectively, which is very similar to the $b=-8$
side.  Also, like the $b=-8$ side, there is marginal evidence in 
Figure~\ref{fig-lon7} that the foreground and background peaks
are closer to the Galactic center at $l<0$ and $l>+4$ respectively.

\begin{figure}[ht]
\centering
\includegraphics[angle=0,width=8cm]{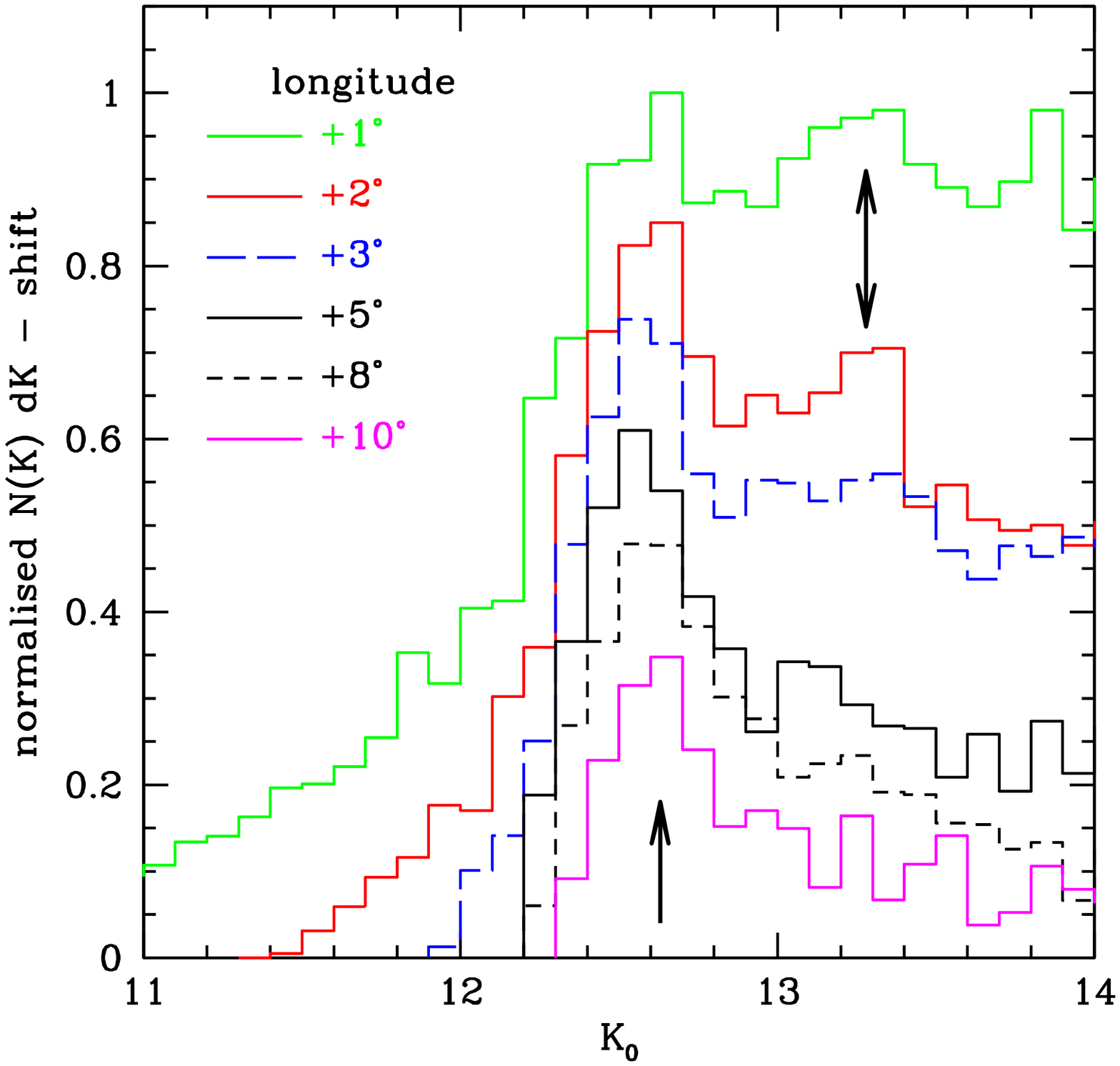}
\caption{Luminosity functions for positive longitudes ranging
from $l$=$+$1 to $+$10$^{\circ}$ at latitude $b$=$+$8, normalized to
the maximum of the bright RC and shifted for clarity.  The position of
the bright RC appears unchanged over 9$^{\circ}$ in longitude, suggesting
roughly constant distance.  The faint RC can also be seen in the profiles
at $l$=$+$1$^{\circ}$ and $+$2$^{\circ}$, and marginally at $+$3$^{\circ}$
longitude.}
\label{fig-over8}
\end{figure}

In Figure~\ref{fig-over8} we overplot the luminosity function at 
$b$=$+$8$^{\circ}$ for longitudes from $+$1 to $+$10$^{\circ}$, 
showing that the foreground RC lies at almost constant
magnitude, although it does seem slightly fainter (more distant) at
$l$=$+$10$^{\circ}$ longitude.  This structure is a major component of
the bulge at positive longitudes; its near-constant distance is completely
at odds with a bar tilted at $\sim$20$^{\circ}$ to the line of sight,
which should have changed distance by more than 2 kpc, over 9$^{\circ}$ in 
longitude, or about 0.7 magnitudes in brightness.

\begin{figure}[ht]
\includegraphics[width=9cm]{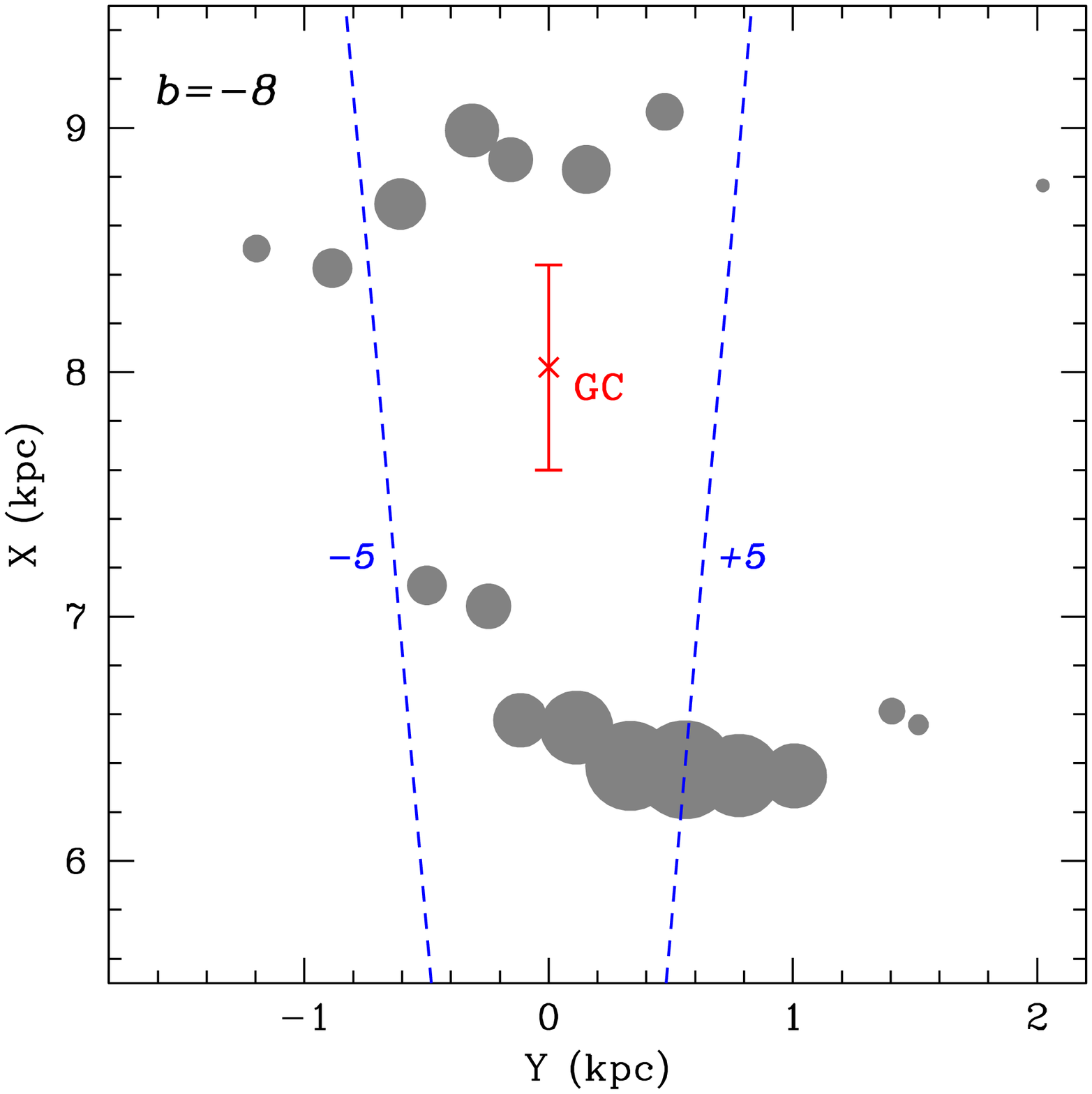}
\caption{Plot of RC distances, in the $b=-8^{\circ}$ plane, for
longitudes $l=-9^{\circ}$ to $+14^{\circ}$, if we interpret
the clumps as tracers of two populations (M$_K$=$-$1.44 assumed
for the $b=-8$ field).  Point sizes are proportional to the number
of RC stars in the peak, 0.1 mag., bin above the background.  
The bright (\/faint) population dominates at positive (\/negative)
longitudes.  Approximate uncertainty on these distances is 
$\sim$0.3 kpc.  The ``GC'' point marks the distance of the central
super-massive black hole (i.e. for $b=0$) from Ghez et al. (2008), 
at 8.0$\pm$0.6 kpc, but projected onto the $b=-8$ plane, for
a distance of 8.1 kpc.
}
\label{distdot-8}
\end{figure}

In Figure~\ref{distdot-8} we plot the RC distances for 12
fields in the latitude $b=-8^{\circ}$ plane, using the RC peak magnitudes
measured from Figure~\ref{fig-lon}, supplemented with four extra fields.
The distances assumed [Fe/H]=$-$0.18 for $b=-8$ with a metallicity correction
of $-$0.20 mag. applied to the observed 47 Tuc RC M$_K$ (see Koch \& McWilliam 2008),
based on the theoretical RC calibration of Pietrinferni et al. (2004, 2006).
The fields range in longitude from $l=-9^{\circ}$ to $+13^{\circ}$.
The sizes of the dots in Figure~\ref{distdot-8} are proportional to the
number of stars at the peak of the RC.

From Figures~\ref{fig-lon} and \ref{distdot-8}, our best
estimate of the maximum counts of the foreground and background RC
populations are $l=+5\pm1^{\circ}$ and $l=-2.5\pm1^{\circ}$ respectively.
A line connecting these background and foreground peaks is tilted to the
$l=0$ line by $\sim$20$\pm4^{\circ}$, which is identical to the tilt
of the Galactic bar claimed by numerous studies.  The line joining the
foreground and background RC peaks intersects the $l=0$ line at a distance
of $\sim$7.7 kpc from the Sun, within 1$\sigma$ of the distance to the
Galactic center (Ghez et al. 2008), projected onto
the $b=-8$ plane.  The $l=0$ intersection with this putative bar line is close
to the half-way distance between the foreground and background, indicating 
symmetry.

The above facts suggest that the RC populations we have found reside at the
ends of the Galactic bar; presumably, at $b=\pm8^{\circ}$ we are seeing 
vertical projections from the bar ends.  The almost constant distances
found for the foreground RC population in Figure~\ref{fig-over8} most
likely reflect stars from the nearby end of the bar, spread out into an
orbital arc, or partial arc.  Our observation that far from the bar ends the
RCs appear closer to the Galactic center distance suggests 
arcs extending from the bar ends; they seem closer to the center than
expected from a circular orbit.  Because the bar is nearly pointing toward us,
the arcs extend roughly perpendicular to the line of sight direction;
this can explain why the distance to the foreground RC does not change much over
9$^{\circ}$ in longitude.

Although several groups have modeled the Galactic bar using the red clump
magnitude as a measure  distance, almost all used low-latitude data, with
$|b|\leq4^\circ$.  
The exception is Rattenbury et  al.  (2007), who included data from 45 OGLE-II
fields; only two of these fields exceeded $|b|=4^{\circ}$, and were
located near $(l,b)\approx(0,-6)$.

Rattenbury et al. (2007) found that beyond longitude  $|l|$$\sim$6$^{\circ}$ 
the  V$-$I luminosity  functions are inconsistent  with their  initial 
simplistic tilted  bar  model used to fit the inner regions.  At  large
negative longitudes the I$_{\rm V-I}$ luminosity functions on the right panel
of Figure~7 in Rattenbury et al. (2007) appear to have the same peak
value, consistent with no tilt to the bar at all, although they did not
comment on this.  For large positive
longitudes,  {\it l}$>$6$^{\circ}$  the OGLE  V$-$I luminosity  functions in
Figure~7  of Rattenbury  et  al.  (2007) were  much  broader than  the
negative longitude  data with a  shorter distance modulus.   For these
OGLE  fields (numbers  8--13)  the bar  model  predicted much  shorter
distances  than the  data.  To  fix these inconsistencies  Rattenbury et
al. (2007)  introduced a  tri-axial model for  the bar, but  this added
complexity still did not provide satisfactory fits to the luminosity
functions at large longitudes positive or negative.

The CMD the ($l$,$b$)=(0,$+$1) field of 
Babusiaux \& Gilmore (2005), published in Figure~4 and 5 of their paper,
indicate a double, tilted, RC separated by 0.7 mag. in the K-band.  The two
RCs show the same (J$-$K$_s$) range and upward tilts indicating that the
(J$-$K$_s$) range is dominated by metallicity dispersion, rather than reddening
dispersion.  While this apparently confirms our double RC finding to $b$=$+$1,
the separation is larger than expected from Figure~\ref{fig-lat} in this work.
They also found a second RC at l=$+$5, indicating a more distant structure.

Nishiyama  et al.  (2005) noted  the presence  of a  weaker RC  in {\it
b}=$+$1 fields with $|l|<7^{\circ}$, at K$_{H-K}$$\simeq$ 13.5.  They also
found a  significantly shallower  slope  of dereddened  K magnitude  versus
longitude  in within $|l|\simeq$$-$4,  compared to  larger longitudes,
and  suggested that this  is evidence  of an  inner bar.   

A  further  confirmation  of  the  existence  of  a  double  clump  at
$(l,b)\approx(0,-8)$ comes from the  analysis of Vieira et al. (2007), 
who  measured  proper  motions  for  $\sim 21,000$  stars  in  Plaut's
Window. The double clump is barely visible in the CMD of their Figure~4,
but it  is evident in their  Figure~10, although they did not comment on
it.  The separation  between the two
clumps  is $\Delta K\sim  0.6$ mag,  consistent with  what we  find in
Figure~\ref{fig-lon},  and slightly larger  than the  separation between
the clumps at $b=-6^\circ$, as if the two populations get further away
from each other when going far from the galactic plane.

\subsection{Extent in Latitude}

\begin{figure*}[ht]
\centering
\includegraphics[angle=0,width=15cm]{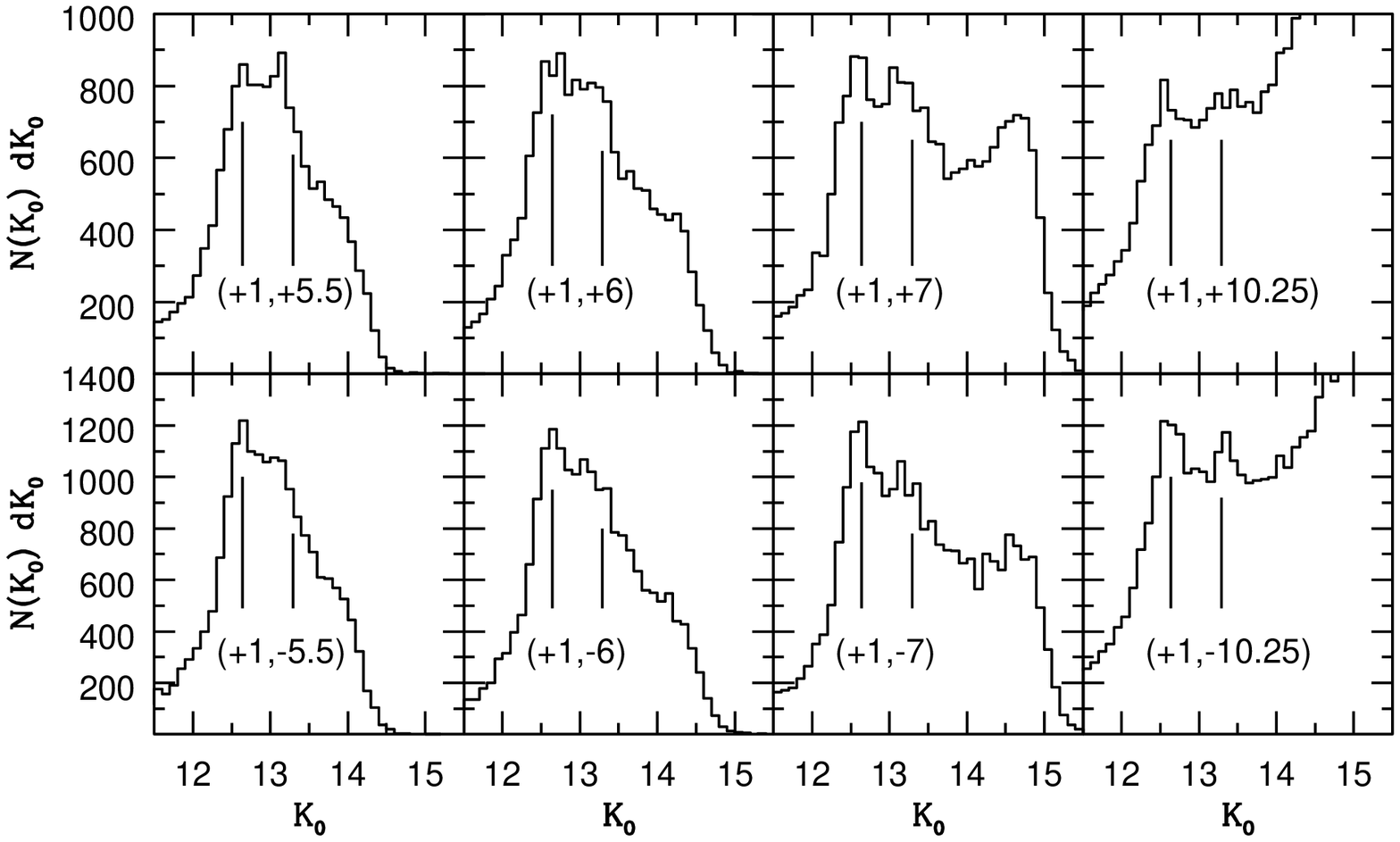}
\caption{RC K$_{0}$ luminosity histograms at l=$+$1 for various
latitudes.  Both bulge RC components are visible from $+$10 to $-$10 degrees latitude.
Toward the Galactic plane the two components merge together.
Tick marks indicate the RC peak positions, from ($l$,$b$)=($=$1,$-$8) in 
Figure~\ref{fig-lon}.  For all except $l$=$\pm$10.25 2MASS stars are from a 30 arc
minute radius circle on the sky.  We have scaled numbers in several panels for
clarity.  Scale factors on the top row are 1.0, 1.09 and 2.0
for panels at $l$=$+$5.5, $+$6, and $+$7 respectively.  On the bottom row
scale factors are 1.0, 1.13, and 1.83 for $l$=$-$5.5, $-$6, and $-$7.  For
$l$=$\pm$10.25 there is no scaling, but 2MASS stars were taken from boxes
1.5$\times$4.0 degrees on the sky for sufficient signal to identify the RC.
}
\label{fig-lat}
\end{figure*}

In  Figure~\ref{fig-lat} we show  a vertical  cut in  the 2MASS  data at
constant longitude, $l=+1^\circ$, with latitudes ranging from $+10.25$
to $-10.25$  degrees.  The main point to notice in Figure~\ref{fig-lat}
is that the  two RC populations are present over a 20 degree range in
latitude, although  we do  not have information  for the  inner $\pm$5
degrees.  We note that data  for the panels at $b$=$\pm$10.25 degrees were
taken from  rectangular boxes,  1.5 degrees in  latitude by  4 degrees
longitude,  for  an  increased  area  of  sky, necessary to detect the
two  RC components  above  the  background   noise.   Thus,  these  panels  in
Figure~\ref{fig-lat}  should  be  divided by a factor of 7.64 to
normalize  to  the  counts   of  the  lower  latitude  fields  (30
arc-minute radius  circles).  Figure~\ref{fig-lat} (and  the $+$1,$-$8
panel  in Figure~\ref{fig-lon})  show that  the  separation between the
foreground and background RCs decreases toward the Galactic plane.
At $|b|$$\sim$$\pm$10$^{\circ}$ the separation in K$_0$ is 0.71 mag., at
$b$=$-$8$^{\circ}$ it is 0.64 mag., while at $|b|$=$\pm$5.5$^{\circ}$
the separation is 0.44 mag.  This indicates that the two RCs are closer
together at lower Galactic latitudes.
This is  consistent with the  de-reddened OGLE V$-$I photometry of Baade's
Window  (at $b$=$-$4) in Figure~1, showing a single, wide  RC.

In Figure~\ref{fig-lat} the bright RC appears approximately unchanged
with latitude, while the faint RC becomes brighter toward the Galactic plane.
This suggests that the bright and faint RCs are closer together in distance
at lower latitudes.  Because the K$_0$ value of the bright RC does not change
significantly it might be assumed that the bright RC is at a fixed distance,
and that the faint peak is closer at lower latitudes, forming a K-shape
morphology.
However, for Figure~\ref{fig-lat} the computation of the de-reddened K-band
magnitudes employed a single RC M$_K$ value, which we believe is correct
for the metallicity of the field at $b$=$-$8$^{\circ}$, but is not correct 
for fields with different mean metallicities.  

\begin{figure}[ht]
\centering
\includegraphics[angle=0,width=9cm]{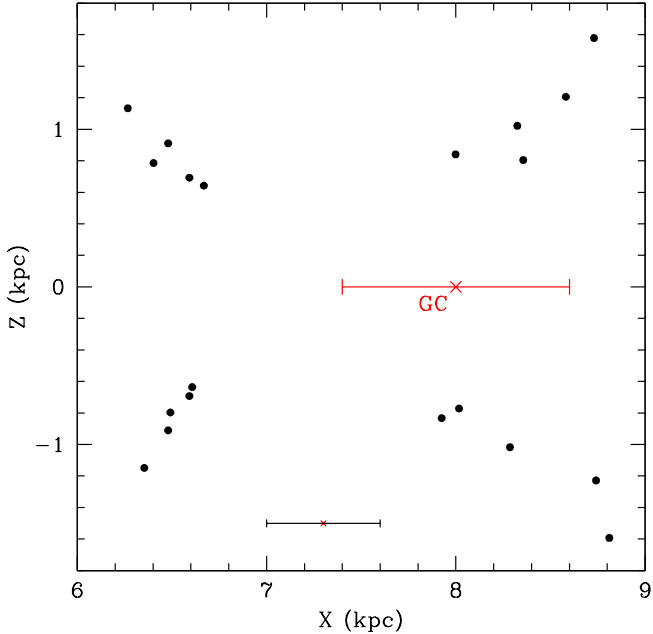}
\caption{Distances of the two RCs as a function of latitude (i.e., in the X-Z
plane), computed using 47~Tuc RC M$_K$ corrected for 
Zoccali et al. (2008) mean [Fe/H] values, based on Teramo group theoretical RC
M$_K$ predictions (Pietrinferni et al. 2004).
A dominant, nearly symmetric, X-shape morphology is obvious.
The ``GC'' point marks the geometric Galactic center distance determined by
Ghez et al. (2008), at 8.0 $\pm$0.6 Kpc.  The X-shape appears centered
at 7.3 $\pm$0.3 Kpc; thus, to within the measurement uncertainties, the X-shape
center agrees to within the uncertainties with the Galactic center distance.
However, better agreement would occur if the bulge is 2--3 Gyr younger than
47~Tuc
}
\label{fig-x}
\end{figure}

In order to determine more reliable distances from the two RCs we adopt
metallicities as a function of Galactic latitude from Zoccali et al. (2008),
and the metallicity sensitivity of RC M$_K$ from the Teramo stellar evolution
models (e.g., Pietrinferni et al. 2004).  We note that the RC M$_K$ is also sensitive
to age;  in particular, the Teramo RC models show steeper metallicity-dependence
for older ages.  Studies of the Galactic bulge age (e.g., Zoccali et al. 2003;
Clarkson et al. 2008) show that it is in the range 10--14 Gyr, with a very small
population $\leq$5\% possibly as young as 5 Gyr.  We have adopted an age
of 12 Gyr and assume a 1$\sigma$ uncertainty of$\sim$1 Gyr.  We simply add the
theoretical RC corrections to the observed M$_K$ of the 47~Tuc RC.
In Figure~\ref{fig-x} we show the distances of each RC for each latitude field
in Figure~\ref{fig-lat}.  Figure~\ref{fig-x} shows a remarkable X-shape structure;
again, we note that if the metallicity gradient were not taken into account 
Figure~\ref{fig-x} would appear more K-shaped.

An alternative interpretation is that the metallicity of the background field
increases toward the Galactic plane, while the foreground field remains at
constant mean metallicity.  This explanation requires a metallicity increase of
$\sim$0.7 dex from $|b|$=10.25 to 5.5$^{\circ}$, which is much greater than
the observed metallicity gradient of Zoccali et al. (2008).  We therefore
abandon this possibility.

Recently, Zoccali et al. (2010), Hill et al. (2010, unpublished) and
Babusiaux et al. (2010) claimed that
the bulge metallicity gradient of Zoccali et al. (2008) is an artifact,
due to a changing ratio of two populations of bulge stars with different
metallicities and kinematics.  The larger velocity dispersion bulge stars have
lower mean metallicities (e.g., Minniti 1996), and
reside further from the Galactic plane than the higher metallicity population.
We note that the range of [Fe/H] from Zoccali et al. (2008) is such that
almost all of the Helium-burning stars in both populations go through the RC phase.
Our use of the mean metallicities is appropriate to determine average distances
as a function of longitude and latitude, but it is probable that the two populations
do not share the same morphology.

We note that there must also be a third population, the inner-halo, as evidenced
by the bulge RR Lyrae stars; this population possesses a spherical, rather
than bar-like, morphology (Majaess 2010).  However, the inner halo 
does not affect our results because most of its stars would not appear on the
RC and because the peak of the halo metallicity function, near [Fe/H]=$-$1.6 dex,
was not detected in the study of Zoccali et al. (2008).

Thus, our picture of the 2MASS bulge now suggests a latitudinal narrowing of the
two RC populations toward the Galactic center, with roughly constant distances
in the longitudinal direction at high latitudes, together with an asymmetry
in longitude.  This suggests a three dimensional X-structure.

We estimate the minimum fraction of RC stars in the peaks in Figure~\ref{fig-lon} by
subtracting the RGB background below the RC peaks, through interpolation of
the RGB above and below the RC region.  The minimum counts between the two
RGB background-subtracted RC peaks provides a means to estimate the minimum
contribution of the peaks to the total RC population.  Either the minimum is due
to the overlapping wings of the peak profiles, or a smooth RC population, due to
the bar, underlying the RC peaks.

Between 20 and 100\% of the bulge RC at ($l,b$)=($+1,-8$) is in either the
foreground or background peaks.  At ($l,b$)=($-1,-9$) the minimum extends practically to
the background, so the entire population appears to be either background or
foreground at this latitude, with no detectable bar at intermediate distances.
For latitudes with $|b|<6$ (c.f. Figure~\ref{fig-lat}) it is difficult to determine
whether the foreground/background components are a few percent or 100\%
of the population.  Detailed modelling will be required for more precise estimates.

The narrowing of the distance between the foreground and background RC populations
suggests that the bar is significantly shorter than the distance between the two
RCs at $|b|=8$.  It is also possible that the X-shape dominates over the bar
even close to the Galactic plane; however, the BRAVA velocity dispersion is consistent
with a bar component, even at $b=-8$.

\section{Other Differences Between the Two Populations}

Here we investigate published evidence relevant to
differences between the bright and faint RC populations.

\subsection{Proper Motions}

If our bright and faint RC bulge populations are on opposite sides of
the Galactic center, then their different distances can be revealed by proper
motion studies.  
Mao \& Paczy\'nski (2002) predicted a proper motion difference between bright
and faint RC bulge sub-populations near 1.6 mas yr$^{-1}$ for Baade's Window,
assuming that they correspond to the near and far halves of the Galactic 
bar.

Sumi et al. (2003) measured proper motions of 47,000 bulge stars from OGLE 
survey data (at $l,b=+1,-3.6$), with a baseline of 4 years.  
They found a proper motion difference of 1.5$\pm$0.06 mas yr$^{-1}$ between
bright and faint RC sub-populations, in good agreement with the expectations of 
Mao \& Paczy\'nski (2002).  Our analysis of the Sumi et al. (2003) data
shows that the proper motion difference is reduced to 1.0$\pm$0.06 mas yr$^{-1}$
if the bright end of the bright RC population box is reduced to resemble our
bright RC sub-population limits.  This result indicates that the bright
and faint RC populations are, indeed, separated in distance as we assume here
for our fields at higher latitudes.  

An additional proper motion test can be obtained from the photographic data of 
Vieira et al. (2007, henceforth V07) for the Plaut field at $(l,b)=0,-8$, with 
a baseline of 21 years.  From 328 bright RC and 365 faint RC stars in the V07 
data, with 2MASS photometry, we find only a 1$\sigma$ difference between the
mean proper motions of bright and faint RCs, in the longitude direction,
at 0.19$\pm 0.19$ mas
yr$^{-1}$.  In the Galactic latitude direction the proper motion difference
is 0.51$\pm 0.18$ mas 
yr$^{-1}$, or about 3$\sigma$, with the faint RC mean proper motion smaller than
the bright RC's.  

Our analysis of the V07 data also shows a 3.4$\sigma$ difference in the proper
motion dispersions of the bright and faint RC populations in the latitude direction.
The faint RC dispersion is smaller than for the bright RC, consistent with the faint
RC being more distant, although not as large as would be expected from our adopted
distances.  However, we expect that the signal will be diluted by
contamination from foreground/background RGB stars in the faint/bright RC samples.

Thus, while the V07 mean longitudinal proper motion differences between the
two RC populations are not consistent with distance separation, both the mean
proper motion and the proper motion dispersion in the latitude direction are
consistent with the faint RC at a significantly greater distance.


\subsection{Radial Velocities}


Two radial velocity studies of the Galactic bulge have been undertaken recently:
the extensive BRAVA fiber survey (Howard et al. 2008, 2009) of bright bulge giants, 
and a Fabry-Perot investigation of bulge giants and RC stars by
Rangwala, Williams \& Stanek (2009, henceforth RWS09).  The two studies show
consistent mean velocities and velocity dispersions in the overlap regions.  

The BRAVA study is quite extensive, covering a range of 20$^{\circ}$ in longitude,
at latitudes of $b=-4^{\circ}$ and $-8^{\circ}$.  The resultant BRAVA velocities are
consistent with cylindrical rotation with a peak-to-peak 
amplitude of $\sim$150 km/s for fields at $b=-4^{\circ}$ and $-8^{\circ}$, or a 
maximum velocity of $\sim$75 km/s.  Howard et al. (2008, 2009) conclude that
this is consistent with a pseudo-bulge, but not a classical bulge.
Shen et al. (2010) model the BRAVA velocities and find a pure bar morphology with no 
trace of a classical bulge.
Because the BRAVA survey did not reach the faint RC, it contains no information
on the bright versus faint RC kinematics.

A very important point is that both the BRAVA and the RWS09 velocities are
consistent with orbital motion, showing the expected change in sign 
where stars at positive
longitudes, near $l=+5^{\circ}$, are foreground, but background near $l=-5^{\circ}$.  
This strongly confirms our assumption that the fields near these longitudes are 
separated in distance, rather than at the same distance with differing RC magnitudes,
due to unspecified population effects.

One might ask how the BRAVA velocity-longitude plots can appear continuous
and symmetric when it samples stars on opposite sides of the Galactic center
that are asymmetrically distributed in longitude.  The answer is that for a line
of sight through a circular orbit the foreground and background have the same
radial velocity, but opposite proper motions.  The symmetry of the BRAVA
velocities with longitude indicates that the foreground and background populations
are equally distant from the center of rotation.

In addition to the overall velocity field
RSW09 also dissected the RC into bright and faint components. They found 
bright minus faint RC velocity differences in their $l=\pm5^{\circ}$ fields
of $-$35$\pm$11 km/s, which they attributed to non-circular streaming motions.
Closer to the minor axis, at $l=+1^{\circ}$, RSW09 found no velocity differences 
between the two RC populations.

The velocity difference at $\pm$5$^{\circ}$ might be understood as the signature
of off-center bar rotation, with the axis of rotation closer to the background
RC population; however, this possibility appears to be ruled-out by the symmetry
of the BRAVA velocities.
RWS09 noted that this velocity difference is opposite to the 
predictions of Mao \& Paczy\'nski  (2002): faint RC stars have  
more positive velocities, not the expected more negative values.
Thus, the bright/faint RC velocity differences of RWS09 must be investigated
further.

\subsection{Metallicities}

Although many metallicity studies of the Galactic bulge have been undertaken,
only Rangwala \& Williams (2009, henceforth RW09) have compared [Fe/H] values
from bright and faint RC stars.  
RW09 measured metallicities from the 8542\AA\ Ca-triplet line
in their earlier Fabry-Perot data.  Based on a calibration with high resolution [Fe/H]
values they inferred a mean [Fe/H] for the field at $l=+5.5^{\circ}$ of $-$0.55$\pm$0.03 dex,
and $-$0.17$\pm$0.03 dex for the $l=-5^{\circ}$ field.  This metallicity difference
suggests that the bright and faint RC are distinctly different 
populations.  However, such a longitudinal, lop-sided, metallicity difference is very
difficult to understand for most morphologies; if correct, accretion of a dwarf galaxy
seems plausible.

Because a lop-sided metallicity trend would have significant implications for
our understanding of the bulge, independent verification of this result is 
critical.  The presence of stars in the RW09 sample with reported [Fe/H] up to
$+$3 dex gives cause for concern that large systematic uncertainties may be present.
Andreas Koch (2010, private communication) has made provisional
metallicity measurements from the BRAVA spectra; he found no significant
asymmetry with longitude, contradicting the unusual metallicities of RW09.
We, therefore, significantly downgrade the weight given to the metallicity
asymmetry claimed by RW09.

\section{Summary and Discussion}

From 2MASS K$_0$, (J$-$K)$_0$ color-magnitude diagrams we found two
distinct RC populations towards the Galactic bulge, separated by
$\Delta$K$_0$$\sim$0.65 magnitudes at latitude $b=-8^{\circ}$, and co-existing
in fields over 13$^{\circ}$ in
longitude and 20$^{\circ}$ in latitude.  The presence of the two RC populations
is particularly obvious at a Galactic latitude of $-$8$^{\circ}$.
We detect the individual RC populations over a $\sim$20$^{\circ}\times$20$^{\circ}$
area of sky, roughly symmetric about the Galactic center.  Thus, the two
populations cover essentially the entire Galactic bulge/bar region; however,
the 2MASS data do not probe the RC within $\sim$5$^{\circ}$ of the Galactic plane,
due to confusion limitations.
We also find that the faint RC is dominant on the negative longitude
side of the Galactic center, while the bright RC is the principal population
at positive longitudes.

Based on the age and metallicity sensitivity of the RC, predicted by theoretical
stellar isochrones of the Teramo group (Pietrinferni et al. 2004), we find that
the two RCs cannot be due to any allowed combination of age or [Fe/H] in
the Galactic bulge/bar at a single distance.  Number statistics firmly rule-out
the possibility that 
the two luminosity peaks are due to RC plus red giant branch bump (RGB bump) or
asymptotic giant branch bump (AGB bump).  Heretofore unknown stellar evolution
effects are also militated against, due to the change in number ratio between
positive and negative longitude sides of the Galactic bulge.

The obvious, and natural, explanation is that the faint and bright RC peaks
reflect different distances of two populations.  At ($l,b$)=($+1,-8$) we estimate
distances of 6.5 and 8.8$\pm0.2$ kpc for the two RCs, based on M$_K$=$-$1.44,
computed with theoretical offsets from the observed 47~Tuc K-band absolute magnitude,
and assuming an age of 12 Gyr and a mean [Fe/H]=$-$0.18 dex.  Figure~\ref{distdot-8}
and \ref{fig-x} should be consulted for distances at other locations.  We note
that these distances are based on the assumption of a bulge age equal to that
of 47~Tuc, near 12 Gyr; our estimated distances increase for younger bulge ages.

Radial velocity data, particularly from the BRAVA survey (Howard et al. 2008, 2009),
are consistent with the faint RC population being more distant than the bright RC
population.  Proper motion results from Sumi et al. (2003) and V07 are also
generally consistent with a distance interpretation.

An immediate conclusion is that these two co-existing populations are not
consistent with the body of a bar, since the line of sight through a bar should
occur at one distance, not two.  Other evidence against these populations 
being in the body of a bar is the fact that both populations mostly exist at 
fixed distances, independent of longitude; in particular, the foreground RC peak
K$_0$ magnitude is unchanged over 9$^{\circ}$ in longitude.  This is in stark
contrast to a bar, which other studies have claimed is pointed almost directly 
towards us, tilted by only $\sim$20$^{\circ}$ to the line of sight.  For
a bar tilted at 20$^{\circ}$ the distance change over 9$^{\circ}$ in longitude
is expected to exceed 2kpc, or roughly 0.7 magnitudes.

When we plot distance of the peaks versus longitude for latitude $b=-8^{\circ}$,
and add approximate number counts, we find that the faint and bright RC counts
peak near longitudes $l=-2.5^{\circ}$ and $l=+5^{\circ}$, respectively.  A
line joining these two maxima is tilted, approximately 20$^{\circ}$ to the line
of sight and crosses the $l=0$ line within 1$\sigma$ of the distance to the
Galactic center, projected onto the $b=-8^{\circ}$ plane; this intersection point
is roughly midway between the foreground and background populations.

As a function of latitude our luminosity plots show that the two RC peaks are
closer in distance at lower Galactic latitudes (i.e., toward the plane), mostly
because of an increase in brightness of the background RC.  The two RCs appear
to merge around $b=\pm5^{\circ}$.
When we make a crude correction to the RC M$_K$ for the metallicity
gradient with latitude along the minor axis, by Zoccali et al.
(2008), we found that a vertical cut through the bulge/bar, at $l=+1^{\circ}$ (from
$b=-10.25^{\circ}$ to $b=+10.25^{\circ}$, i.e., as viewed from the side) shows an
X-shaped morphology.  The center of the X-shape occurs near 7.3$\pm$0.3 kpc, 
which is equal to, within the uncertainties, the Galactic center distance of
Ghez et al (2008), at 8.0$\pm$0.6 kpc.  
Better agreement between these two centers would be obtained if the bulge is
$\sim$2--3 Gyr younger than 47~Tuc, or if the metallicity difference
between bulge and 47 Tuc RC stars is larger than we assume, or if there is an
unspecified systematic error in the theoretical isochrones.

The narrowing of the distance difference approaching the Galactic plane suggests
that if this X-shaped structure is connected to the bar, then the bar is
significantly shorter closer to the plane.

The RC peak counts above the troughs in the double peaked RC luminosity functions
gave a minimum estimate for the fraction of stars in the X versus stars in a broader
RC distribution, perhaps from a bar or spheroidal component.  This is a minimum
because it is possible that the troughs are
partly, or entirely, due to the overlapping wings of the distributions of the
foreground/background
RC components.
We found that at ($l,b$)=($+1,-8$) the foreground/background RC X components
are at least 20\% of the total RC population; at ($l,b$)=($+1,-9$) the
foreground/background X-components must be close to 100\% of the total RC;
on the other hand, near $b=\pm5.5$ the estimated fraction due to foreground/background
components ranges from almost negligible to 100\% of the population.

Our analysis of the 2MASS data contains no compelling direct evidence of a Galactic
bar, although one may be present in a smooth RC population between the RC peaks.
As noted above, at ($l,b$)=($+1,-9$), the 2MASS data show no room for
a bar.  Detailed modelling will be required to constrain a bar population in this
data.  It seems reasonable, however, that the X-shape may merge into a short
bar at latitudes below $|b|\sim5$.  We note that the radial velocity data of
the BRAVA survey shows an increased dispersion near $|l|=0$, as expected
from a bar.  Detailed models of these radial velocities by 
Shen et al. (2010) provide excellent fits with a pure bar structure; they
found that a spheroidal
component can be no more than 8\% of the total.  Our finding that at least 20\%
of RC stars are in the X-shape component at $b=-8^{\circ}$, rather than a bar,
may be in conflict with the 8\% limit of Shen et al. (2010); however, their
limit refers to a spherical distribution, rather than arcs at the ends of a
bar, so it is possible that no conflict exists.

A reasonable, qualitative, model to explain the observations presented here
is that the high-latitude bulge (above $|b|\sim$6$^{\circ}$) is dominated
by the vertical extent of stars near the foreground and background ends of
a Galactic bar.  In this scenario stars extend in all directions
from the ends of the bar, i.e., in both longitude and latitude.
In longitude the stars occupy arcs, or partial orbits, emanating
from the bar ends.  For this reason, a line connecting the peaks in the foreground/
background RC populations trace the direction of the bar, tilted at $\sim$20$^{\circ}$
to the line of sight, assuming that there is no significant rotational lag.
The axis of rotation of the bar is marked by the point
where this line intersects with $l=0$.  In Figure~\ref{fig-x} this point is located
at 7.7 kpc from the sun; it happens to lie roughly midway between
the two RCs and within 1$\sigma$ of the Galactic center distance.  This arrangement
should produce symmetry in the mean radial velocity curve, similar to the BRAVA
observations.
Because the bar is almost end-on, the stars in the arcs, near
the bar ends, spread almost tangentially to the line of sight, and thus appear
at a nearly constant distance, as observed.
However, at large angular distance, in longitude, from the bar ends, the stars
along the arcs/orbits are closer to the Galactic center distance, also observed.  Since 
stars emanating from the bar ends do so in all directions, they appear as two
roughly circular areas, and take-on a peanut appearance when viewed nearly end-on
to the bar; a slight flattening of the peanut may be expected due to the
gravitational potential in the latitude direction.  When viewed from the side,
the structure takes on an X-shape morphology, as the stars emanating from the
bar ends merge into the bar.  In this case the main component of the bar resides
at low Galactic latitudes, and perhaps for this reason the end components are
relatively easy to detect at high latitude.

An alternative description is that there is only an X-shaped component 
which extends to high latitudes, but without connecting to a bar.  As before,
since we are looking almost end-on to the axis of the X, the foreground and
background appear at nearly constant distances.  Another, alternative, explanation
of the observed two RC components is that we have detected concentrations in the thick
disk near the ends of the bar.  If correct, this scenario might naturally
explain the similarity of the composition of thick disk and bulge red giant
stars found by Alves-Brito et al. (2010).

If our two populations represent extensions from the ends of a bar, then their
velocities should show cylindrical rotation, similar to bar models, but with a 
distribution not considered
by Shen et al. (2010).  One major difficulty is that our 
results suggest a end-to-end bar length of $\sim$2.5 kpc, whereas the
best fit model of Shen et al. (2010) indicates a bar half-length of 4kpc.

X-shape, boxy, and peanut morphologies of extra-galactic bulges
are well known phenomena, e.g., IC 4767, NGC 128, NGC 4469, IC 2531 
(e.g., Whitmore \& Bell 1988; Bureau et al. 2006.)  The X-shapes
are relatively minor components of
extra-galactic bulges, and significant image processing is usually
required to highlight them, in stark contrast to the strong signal
we find here for the Galaxy.

N-body simulations of isolated disk galaxies and the growth of bars 
(e.g., Patsis et al. 2002; Athanassoula 2005) find 
X-shape morphologies
resulting from bar growth.  In particular, the {\bf x1v1} resonance mode of
Patsis et al. (2002) appears very similar to our results for the Galaxy.
If these predictions are correct, the implication is that the bulge X-shape
resulted from bar instabilities, and that the Galaxy
contains a pseudo-bulge, and a bar, rather than a classical bulge.  
X-morphologies also result from mergers, although bars are still
involved: Mihos et al. (1995) performed numerical simulations of satellite accretion
by S0 galaxies, and found that prograde accretion induced the formation of bars,
which subsequently buckled and produced X structures.  Since the formation of
a bar from an isolated disk galaxy relies on interactions between its disk and
halo, the overlap between these two mechanisms is greater than at first
apparent.

Pseudo-bulges and bars are expected to display cylindrical rotation 
(e.g., Athanassoula 2005), which is consistent with the radial velocity BRAVA survey
as noted by Howard et al. (2008, 2009).

Typical timescales for pseudo-bulge growth is $\sim$1 Gyr (e.g., Athanassoula 2008),
which may be problematic for a rapid formation timescale for the Galactic bulge
(e.g., Ballero et al. 2007), as implied by the enhanced [Mg/Fe] ratios
(e.g., Fulbright et al. 2007; Lecureur et al. 2007; McWilliam \& Rich 1994);
although, both the timescale and the Mg enhancements have associated uncertainties
which may bring them into rough agreement.

Another difficulty is that pseudo-bulges/bars are thought to
result in well mixed orbits, such that metallicity gradients may 
not be expected (Zoccali 2010; Howard et al. 2009.) However, the bulge
metallicity gradient is well-known (Zoccali et al. 2008), and the
metallicities correlate with kinematic properties.  It is possible that
the gradient simply results from the super-position of two populations 
(Zoccali et al. 2010; Babusiaux et al. 2010; Hill et al. 2010.)
Howard et al. (2009) speculated that dissipative processes during
bar formation might explain the metallicity gradient.

While an X-shape bulge qualitatively explains the 2MASS photometry
and much of the published kinematic data, some literature results are
problematic.  The $-$35 km/s radial velocity difference
between bright and faint RC found by RWS09 is difficult
to understand; asymmetric bar rotation could explain it, but such an
asymmetry in velocities is not obvious in the BRAVA data.  Also, the
0.4 dex difference in [Fe/H] between $l=+5$ and $-5^{\circ}$ bulge
components, found by RW09, suggests an accreted system,
rather than a bar or symmetric X-component; however, provisional
metallicities from the BRAVA spectra, by Koch (2010, private communication)
do not support an [Fe/H] asymmetry.  Perhaps the most important
difficulty is that the X-shape is evident in our 2MASS data when the
dependence of mean [Fe/H] on Galactic latitude is taken into account,
without the metallicity corrections the morphology is more K-shaped.  This
is a problem because the expectation is that the bar should not possess a
metallicity gradient, and while X-morphologies are well known in external
galaxies, we are unaware of K-morphologies.  
The final piece of confusing evidence is the lack of difference between
the mean proper motions of bright and faint RC stars in the data of 
Vieira et al. (2007).


These apparent inconsistencies demand that future work on the kinematics and
composition of RC bulge stars be pursued for a more complete picture of the
Galactic bulge region.  It is particularly important to verify whether there
is a metallicity gradient with latitude in the RC X-populations; if so, this
would support the idea of Howard et al (2009), that the bar formed with dissipation;
if not, then the bulge is K-shaped.  If the gradient simply reflects a
changing ratio between two populations with different mean [Fe/H], then
composition studies of these two populations would be interesting and useful.
A definitive investigation of the possible metallicity asymmetry with longitude
proposed by Rangwala \& Williams (2009) is also necessary.  Extant age studies by
Zoccali et al. (2003) and Clarkson et al. (2008) should be supplemented with
a photometric survey for the ages of bulge stars at $(l,b)$=$(\pm5,-8)$, in order
to completely rule-out the possibility that age is responsible for the luminosity
difference between the two RCs at this latitude.

Other important future work includes detailed modelling of the 2MASS data, in order
to determine the relative proportions of a bar and the foreground/background RC
components.  A high spatial resolution infrared photometric survey,
such as the VISTA Variable in the V\'ia L\'actea survey 
(VVV, Minniti et al. 2010), covering roughly the same longitudinal extent as
the current work, but going to much
smaller latitudes will help to determine whether the X-shape continues to
the Galactic center, or joins the bar. Analysis of the premilinary VVV 
catalogues is ongoing (Saito et al. 2010, in preparation).
Independent verification of the foreground/background distances,
at high latitude, with standard candles, such as Miras, strong-lined RR Lyrae 
stars, and eclipsing binaries, will also be very helpful.

After this paper was submitted for publication Nataf et al. (2010)
reported a split RC toward the Galactic bulge.  Their findings are broadly
consistent with the results of this paper; in particular, they find very
nearly equal (V$-$I) colors, consistent with similar metallicity for the two
RC components.


\acknowledgements
\centerline{\bf\large Acknowledgements}

Part of this work was performed by AM while visiting the KITP in Santa
Barbara,  which   is  supported   under  NSF  grant   PHO5-51164.  AM thanks
Andreas Koch for a useful conversation.   MZ
acknowledges  the FONDAP  Center for  Astrophysics  15010003, the BASAL
Center for Astrophysics and Associated Technologies PFB-06, FONDECYT 
Regular 1085278, and the MIDEPLAN Milky Way Millennium Nucleus P07-021-F.

\end{document}